\documentstyle[art12]{article}

\begin{document}

\pagestyle{empty}

\noindent
\hfill OCHA-PP-55

\noindent
\hfill April(1995)

\vfill

\begin{center}
  {\LARGE Baryon Asymmetry: }

{\LARGE  Evidence of CP Violation and \\
          Phase Transition in the Early Universe ?\footnote{invited talk given
at 23rd INS Symposium
on Nucleon and Particle Physics with Mesons Beams in the $1 Gev / c $ region}}
\\
        \vfill
        \vspace{0.1in}
   {\large Akio Sugamoto \footnote{email: sugamoto@phys.ocha.ac.jp}}\\
        \vspace{0.1in}
         {\it Department of Physics,\\
           Ochanomizu University \\
           2-1-1, Otsuka, Bunkyo-ku, Tokyo 112, Japan}\\
       \vspace{0.2in}
 \end{center}
 \vfill
\begin{abstract}
A talk was given on the baryon asymmetry of our universe within the electroweak
energy scale which was the theme for the 23rd INS Symposium.
It was intended for  non-experts by a non-expert speaker.
A model is analized explicitly in which the lepton number produced from the
bubble walls is converted afterwards to the baryon asymmetry.
Phase transition dynamics is simulated, including the temporal develpment and
the fusion effect of the nucleated bubbles.
\end{abstract}

\newpage

\pagestyle{plain}
\setcounter{page}{1}

\section{Introduction}

Baryon Asymmetry is the problem to explain why baryons (constituents of the
matter) dominate asymmetrically over anti-baryons (those of the anti-matter) in
our present universe,
$ n_{B} \gg n_{\bar{B}} $,
 and to give the number  $ n_{B} / n_{\gamma} \sim 10^{-8} \sim 10^{-10}$ .
 [$ n_{B}$ ,$ n_{\bar{B}}$ , and $ n_{\gamma} $ are respectively the number
densities of baryon, anti-baryon, and photon in our present universe.]
Let us roughly, very roughly estimate the number;

Baryon number $ \sim 10^{-8} Nucleons / cc$ comming from
\begin{itemize}
\item the average distance between galaxies : $6 Mpc $
\item the number of stras of the sun type including a galaxy : $3 \times
10^{11} $
\item the mass of the sun : $ 2 \times 10^{33} g $
\end{itemize}

Photon number $ \sim 400 photons / cc $ comming from
\begin{itemize}
\item cosmic Back Ground radiation $ 2.7 K $
\end{itemize}

Here the difficult problems on the dark matter and the helium synthesis are of
course ignored.
First recognize that a naive discussion on this problem gives a difficulty.
Naive means that at high temperature we have thermal distribution of $ n_{B} =
n_{\bar{B}} = 3/4 \cdot n_{\gamma}$.
According to the cooling down of the universe,  baryons and anti-baryons
annihilate each other, resulting non-baryonic mesons.
But, at about the temperature  $ T \sim  20 MeV $ the annihilation stops, where
the reaction rate becomes much less than the expansion rate of the universe,
namely, the reaction cannot catch up with the expansion.
The ratio is then freezed, giving $ n_{B} / n_{\gamma} = n_{\bar{B}} /
n_{\gamma} \sim  10^{-17} $  which is, however, too small.
Therefore, something should be added in order to explain the observed data on
the baryon asymmetry.
These are the A.D.Sakharov's conditions ~\cite{ADS} :
These are (1) existence of the baryon number $ B $ violating interactions,
(2) existence of the $CP$ and $C$ violations,
and (3) existence of the thermal non-equilibrium.
(1) raises the number from $ 10^{-17}$ to $ 10^{-10}$,
(2) makes the difference between $B$ and $\bar{B}$,
and (3) supressed the inverse reaction of  the $B$ number producing one.
$C$ is always violated in the weak interaction so that CP violation is more
important.
Now the baryon asymmetry is the evidence of  these three conditions.

In 1978,  M.Yoshimura, S.Weinberg, and other people ~\cite{MYW} invented the
GUT's scenario baryon asymmetry, where  $B$  is supplied by the heavy
$(10^{16}GeV)$ Higgs' decay $ X \to \bar{q} + \bar{q} $ and $ q + l $ ,
CP  violation is given by the complex phase in the Yukawa couplings, and the
thermal non-quilibrium is realized by the heavy particles' decay, the inverse
reaction of which is naturally supressed at the lower temperature.

\section{Electroweak Baryogenesis}

The GUT's baryogenesis is the physics of $ 10^{16} GeV $.
We wishes to go down to the lower energy scale, say $ 100 GeV $ of the
electroweak energy scale, in order to be included in this 23rd INS Symposium.
This is called Electroweak Baryogenesis after the work by Kuzmin, Rubakov, and
Shaposhnikov ~\cite{KRS} ('85).

One of the motivations is of couse the energy scale of $100 GeV$ is the
experimentally familiar place.
The other motivation comes from the SPHALERON given by Manton ~\cite{MAN3}('83)
and Klinkhammer and Manton ~\cite{KMAN4} ('84).
I did a similar work ~\cite{SUGA} ('83), using Nambu's solution ~\cite{NANBU}
of the monopolium.
Important ingredient is the chiral anomaly.
Anomaly means the violation of the charge conservation in the presence of the
topologically non-trivial gauge field configurations.
Without the configurations, $B$ and $L$ conserve exactly.
However, in the presence of a topologically entangled non-trivial gauge
configuration classified by the integer number $N_{Chern-Simons}$ ($N_{C-S}$),
conservation of $B$ and $L$ is violated.
Instead we have
$\{ B \ \ or \ \ L \} - N_{g} \cdot N_{C-S} = $ conserves.
 [ $ N_{g} $ is the number of the generations.]
If  you are the nuclear physisists, let's think of the Skyrmion, where the
nontrivial (entangled) configuration of the meson fields gives the proton
having the baryon number.
Here the Skyrmion-like objects are the vacua having zero energy, which are
classified by the integer numbers of $N_{C-S}$.
Now the Sphaleron is the saddle point solution of the Weinberg-Salam model,
located in between the two different vacua, having the energy of about $ 10 TeV
$ with $ N_{C-S} = 1/2 $.
Therfore the Sphaleron controlls the transition between the two different
vacua.

We have the following chemical reaction between three kinds of "atoms";
\begin{equation}
[B] + [L] + [vacuum, N_{C-S}]  \longleftrightarrow  [B+N_{g}] + [L+N_{g}] +
[vac, N_{g}-1]
\end{equation}
\begin{center} Sphaleron Transition \end{center}
$B$ and $L$ are violated, but keeping $ B-L $.
We can consider the following two cases:

Case 1.
Sphaleron transition rate $ \gg $ expansion rate.
Then, the thermal equilibrium is realized, where the equilibrium value is
determined by the conserved
$ B - L $  as $ < B > = O(1) \cdot < B-L > $ .
If $ < B - L > \ne 0 $ , then $ < B > \ne 0 $, but if $ < B-L > = 0 $, then $ <
B > = 0 $ .
The former mechanism is originally adopted by Fukugita and Yanagida7
{}~\cite{FY}('86), and is used in the unbroken phase of the model in the next
section.
The latter is the sphaleron's washing away mechanism.

Case 2.
Sphaleron transition rate $ \gg $ expansion rate $ \star $ .

In this case, thermal non-equilibrium is realized and we have a possibility of
having $ < B > \ne  0$ .
But the condition $ \star $ gives a severe constraint of  $ m_{H_0} < 45 GeV $,
 compared with the LEP data of $ m_{H_0} > 58 GeV $.
We can, however, increase the upper bound by introducing additional bosons.
Introduction of the additional singlet Higgs scalar increases the bound up to
$150 GeV$ due to Anderson and Hall ~\cite{ADH} ('92).
We will use this mechanism, which is just the thing wanted, in the broken phase
of our model.
[ The additional Higgs doublet may raise the upper bound to $190 GeV$.]

\section{The Model}
Now, let us examine the model presented by A.G.Cohen, D.B.Kaplan, and
A.E.Nelson ~\cite{CKN}, based on our work ~\cite{SAY} performed in
collaboration with my student Azusa Yamaguchi.
The model is the standard model modified by the see-saw mechanism
{}~\cite{seesaw} with the additional singlet scalar $\phi$ and the right-handed
neutrinos ${N_{R}}$.
The vacuum expectation value $ <\phi > \ne 0 $ violates the  L-conservation
spontaneously.
This $ L \ne 0 $ introduces the $ (B-L) \ne 0 $ which is converted to $ B \ne
0$  by the fast sphaleron transition of the Case 1 in the unbroken phase where
the recovered  L-conservation protects the washing away of the produced $ L $
(or $ B - L $ ) .
The Lagrangian reads
\begin{equation}
{\cal L}=-{\cal L}(standard \ \ model)+ { \nu_L , N_R kinetic
terms } +  \bar{\Psi(x)^{\cal c}} M(x) \Psi(x) \label{eq:Lag}
\end{equation}
where
\begin{equation}
\Psi(x)=\bigl[ \nu_1, \nu_2, \cdots ,\nu_G, N_1, N_2, \cdots
,N_G \bigr]^{T} ,  \label{eq:psi}
\end{equation}
In Eq.(\ref{eq:Lag}) the mass matrix $M(x)$ is given by
\begin{equation}
\  M(x) = \pmatrix{
	      0, & \lambda_D\varphi(x) \cr
      \lambda^{T}_{D}\varphi(x), & \lambda_{M}\varphi(x)^\ast
	  \cr
	  }
	  \  \label{eq:mass},
\end{equation}
where the position dependency of th mass matrix $ M(x)$ comes from the bubble
nucleation in the electroweak phase transition which is of the 1st order (?) at
least in the perturbative analysis.
The phenomenon is similar to the formation of  liquid droplets in the vapor
vessel when the temperature is lowered to a certain critical value $ T $.
Inside the bubble the mass matrix takes the larger value which plays the role
of the potential barrior for the incoming  neutrinos $ \nu_i $ and the
anti-neutrinos $ \bar{\nu_i} $ .

The reflection coefficients $ R $ and  $ \bar{R} $  for the above two processes

$\nu_i \rightarrow  \bar{\nu_j} [\Delta L = -2] $

and
$\bar{nu_i} \rightarrow  \nu_j   [\Delta L = 2] $

can be expressed respectively by
\begin{equation}
Rji = - U^{T}_{jm} D_m (E) U_{mi}
\end{equation}
and
\begin{equation}
\bar{Rji} = U^{\dagger}_{jm} D_{m}(E) U^{\ast}_{mi}
\end{equation}
where $ UM_{broken \ \  phase} U^T = diagonal $, and the analytic expression
for $D_m (E) $  is obtained.
The $L$-production rate $ D_{ji} $  is now obtained by
\begin{equation}
\Delta_{ji} = | R_{ji} |^2 - \bar{|  R_{ji} |}^2 = -2 \sum_{k \neq l}
Im(D_{k}D_{l}^{\ast})\times J^{lk}_{ji}  \label{eq:delij} ,
\end{equation}
with
\begin{equation}
J^{lk}_{ji} \equiv Im (U_{kj}U_{ki}U_{lj}^{\ast}U^{\ast}_{li})
\end{equation}
which is the product of the two complex phases, one from the scattering phase
shift and the other from the $ CP $ phase, $ J $ , expressed similarly as in
the Jarlskog's parameter in the Kobayashi-Maskawa model.
In our case $ J $ can be non-vanishing when $ Ng \ge 2 $ .

Here another difficulty comes out.
Since the universe is so democratic to all the particles, they are in the
common thermal distribution in the broken phase, where they are equally
massless.
In this situation, summation of $ D_{ji} $ over the initial $i$ or the final
$j$ leads to the no $ L$ number production.
This is the  CPT  theorem or the  GIM-like cancellation mechanism.
To avoid this difficulty we introduce the thermal mass $ M(T) $ proportional to
the $ T $ , following  Farrar and Shaposhnikov ~\cite{Fas} ('94).

Thermal averaging of the $L$-flux produced from the moving wall is
approximately given by
\begin{equation}
f_L / T^3 \sim (A \ln\gamma_\omega +  B -  C\gamma_\omega  ) \cdot J ,
\end{equation}

where $ A $ , $ B $ , and $ C $ are $ O( 10^{-3} )$  for an example having
2-generation  n's  with the masses
$ M_1 (T) = T $ and $ M_2 (T) = 0.5 T $ for $  T = 100 $ or $ 200 GeV $.

Here we should notice that the  $ f_L $ depends on the wall velocity $v_\omega
$ ( its $ \gamma $ factor is  $\gamma _\omega $.) .

\section{ The Phase Transition Dynamics}

If  the wall velocity  $v_\omega$   is constant, then the total  $L$ number
produced reads
\begin{equation}
N_L =      f_L ( v_\omega )  v_{\omega}^{-1} \cdot v_\omega  A(t)  dt  ,
\end{equation}
 and the  Lnumber density is
\begin{equation}
  n_L =   f_L ( v_\omega  )  v_{\omega}^{-1} .
\end{equation}
However, $ v_\omega $  is the time-dependent :
\begin{equation}
 v_{\omega}(t) = \frac{dR(T)}{dt} = 2 \Gamma \left(
                              \frac{1}{R_c}-\frac{1}{R(t)} \right)
\end{equation}
where $R_{c}$ is the critical radius with which the bubble is nucleated. The
$\Gamma^{-1}
$ is the friction coefficient $ O(T) $.

Furethermore, the fusion effect of  bubbles occur during the development of the
1st order phase transition.
Like the cooling down of the vapor (unbroken vacuum of the electroweak theory),
liquid droplets of water (bubbles of the broken vacuum) are nucleated, they
fuse with themselves, and finally the whole vessel (the whole universe) is
filled up with the water (broken phase):
We need to know the temporal development of  the total area of the bubble walls
from which the  L  number is produced.
It is incredible to know that for such a difficult problem the theory exists,
which is called the Kolmogorov-Avrami theory~\cite{KA}, within  the restriction
of the critical radius  $R_c = 0 $,the wall velocity $ v_\omega  = const.$ ,
and the nucleation rate $ I = const$.
This restriction should be modified realistically.
About the critical radius ( the minimum radius of the producd bubble, being
obtained from the balancing between the surface energy $\sim +R^2 $ and the
volume energy $\sim -\epsilon R^3 $),
 the latent heat  $\epsilon$( the difference of the energy inside the broken
phase from the one outside the unbroken phase ),
 and the nucleation rate $I$ ( the probabity for the small bubble to overcome
the surface tension ) can be understood from the following:
1-loop effective potential at $T$
\begin{equation}
V = \frac{\lambda_{T}}{4}\phi^{4} - ET\phi^3 + D(T^2 - T^2_{0})\phi^2
\end{equation}
with
\begin{eqnarray}
D &=& \frac{1}{4v^2} (2m_{W}^2 + m_{Z}^2 + 2m_{t}^2 ) \\
E &=& \frac{1}{\sqrt{2} \pi v^3}(2m_{W}^3 + 3m_{Z}^3) \\
T_{0} &\sim& \frac{1}{2\sqrt{D}}m_{H}
\end{eqnarray}
and
\begin{equation}
\lambda_{T} \sim \lambda \sim \frac{1}{2}(m_{H}/v)^2.
\end{equation}

Here we encounter another difficulty.
What is the phase transition temperature $ T $ ?
It may be a little lower than the critical temperature $ T_{c} $ where the
latent heat begins to be non-vanishing ; $ T = T_{c} - \Delta $ .
The value is roughly the mass of the Higgs scalar $ m_{H}$  ($ 100$ or $ 200
GeV$ ? ).
In our problem the time scale of the phase transition is $ 10^{-26} s $ since
the every parameter involved is the weak scale of $ O(100 GeV) $, whereas the
time scale of the expansion rate at the time is $ 10^{-12} s$ .

Therefore
\begin{center}
\begin{tabular}{lcr}
[the time scale of the phase transition] & & \\
\end{tabular}

\begin{tabular}{lcr}
 &  $ \gg $ & \\
\end{tabular}

\begin{tabular}{lcr}
 & & [the time scale of the expansion rate]. \\
\end{tabular}
\end{center}
This means the slowly cooling down (the annealing but not the quenching) of the
universe, during which the phase transition undergoes.
In order to answer the value $\Delta$   we should couple the phase transition
dynamics with the gravity which is responsible for the cooling down of the
universe.
[In this respect we are reminded of the Maxwell's equal area low in the
gas-liquid phase transition. ]

\section{The Simulation}

We performed the simulation10 using  the KEK and the INS computers, including
the time-dependency of the wall velocity as well as the fusion effect of the
bubbles.
At a proper choice of $ \Delta $ for $ 100GeV$ and $ 200GeV $ , we have the
following figures:

Now the total $ L $ number density $ n_L $ can be simulated by
\begin{equation}
n_L  =  \sum_{i} \int f_{L} ( v_{w}^i )  A(t)^{i} dt / V ,
\end{equation}
where the summation is carried out over the various segments $ i $ of the
bubble walls behaving differently.The resut is

 \[ n_L / T^3 = \left\{
	   \begin{array}{@{\,}cccc}
 ours                          &  & Kolmogorov-Avrami (K-A) &    \cr
 -0.299  \times 10^{-2} \cdot J & \leftrightarrow & 0.108 \times 10^{-2} \cdot
J &  ( T_c = 100 GeV ) \cr
  0.303 \times  10^{-2} \cdot J &\leftrightarrow &  0.209 \times 10^{-2} \cdot
J &  ( T_c = 200 GeV ) ,\cr
\end{array}
\right. \]
so that we have
\begin{equation}
n_L (ours)  /  n_L (K-A)  =    2.77/1.45
\end{equation}

or the difference of the factor $ 2 \sim 3 $   occurs depending the details of
the phase transition dynamics.
Here the models adopted are the  2 n's models given above.

\section{Baryogenesis from Leptogenesis}
Chemical equilibrium is used to generate the $ B $ from the produced $ L $ from
the bubblewall.
This is realzed in the unbroken phase (outside of the bubbles) where the
sphaleron transition is very rapid (Case 1 of the Sec. 2), but the $ L $ number
conservation is recovered in this spontaneously broken $ L $-conservation
model.
After $ B $  comes into the broken phase (inside of the bubbles which fill up
the whole universe in the end of the phase transition), $ B $  survives against
the washing out mechanism by the sphaleron, since in this region the additional
singlet scalar supresses the sphaleron's effect (Case 2 of the Sec. 2).
To reproduce the observed value of the baryon asymmetry, $ CP $ violation
factor $ J $ should be $ O( 10^{-5} \sim  10^{-7} )$ .

\section{Conclusion}

\begin{enumerate}
\item In the problem of the electroweak baryogenesis the severe restriction of
  $ m_{H_0} < 45 GeV $ may be avoided with the model of the explicit production
of $ B - L $, where the spontaneous  $L$ violation by the singlet scalar is
essential.

\item  Simulation of the 1st order phase transition is possible by including
the temporal dependency of the bubble-wall's velocity as well as the fusion
effect of the bubbles.
By these effects, the total  $B$  number produced increases by the factor $ 2
\sim 3 $ from the simple model of Kolmogorov and Avrami.

\item A lot of difficulties, however, exist on the following points;
 \begin{itemize}
   \item avoidance of the  CPT by the introduction of the finite $ T $  masses?
;
   \item phase transition temperature ? ,

   phase transition dynamics including gravity? ;
   \item friction? ,

   effective potential or effective action?,

   sphaleron transition? .
 \end{itemize}
\item  How about the reliabilituy of the model and the predicted number?
So far so good, but we are still in the middle of producing various models and
examining them carefully.
It is, however, true that the  CP violation really exists as well as the
thermal non-equilibrium does.

\item The problem of the electroweak baryogenesis includes a variety of various
regions of physics, so that I think it is the interesting problem to pursue.

\end{enumerate}
This is the end of my talk.  Thank you.

\begin{flushleft}

\end{flushleft}
\begin{figure}[t]
\vspace*{10cm}
\caption[nucleation rate]{nucleation rate; $V$ is $T^{4} [GeV^{4}]$}
\label{nuclrate}
\end{figure}%
\begin{figure}[b]
\vspace*{10cm}
\caption[simulation result s1a]{The area of the wall for simulation and
 Kolmgorov-Avrami
(solid line: simulation, dots line: Kolmgorov-Avrami)}
\label{rst1}
\end{figure}%
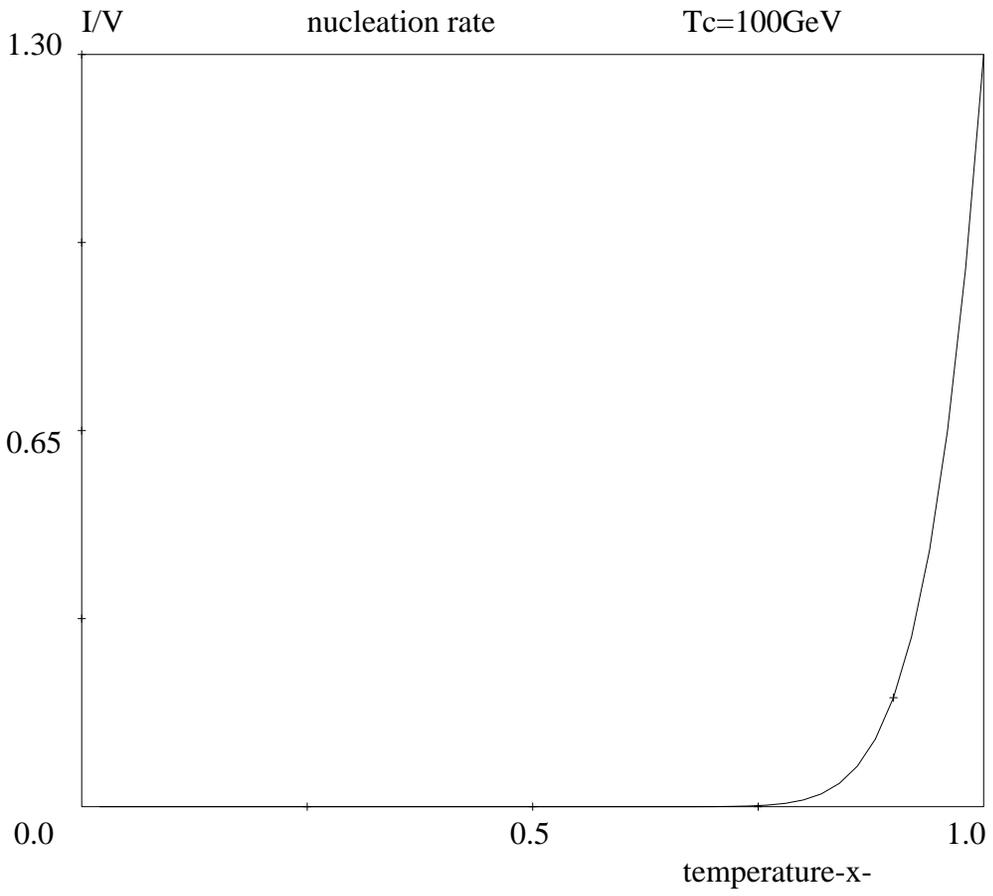
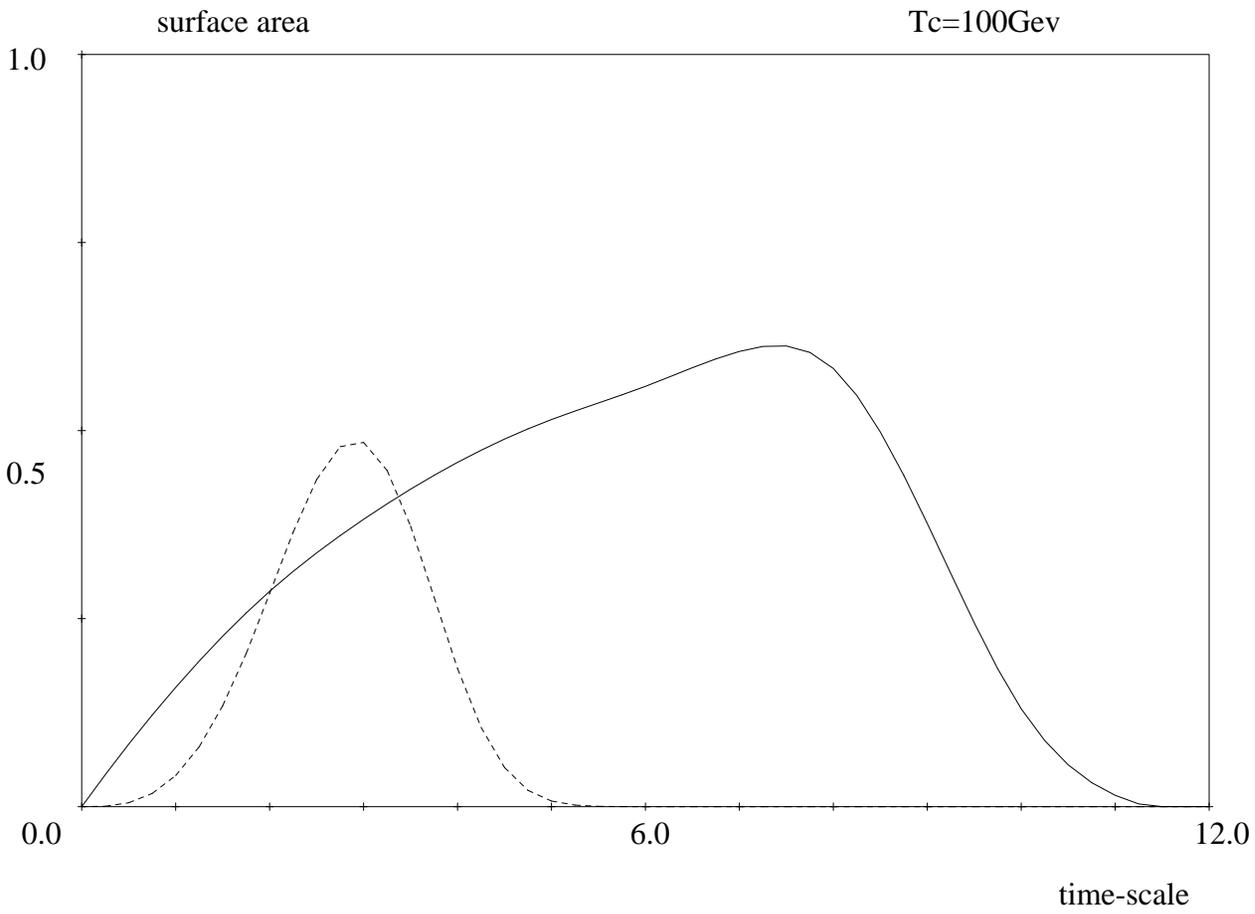
\end{document}